\documentclass[superscriptaddress,twocolumn]{revtex4}

\usepackage{amsmath, amssymb, hyperref, slashed}
\newcommand{\be}{\begin{equation}}
\newcommand{\ee}{\end{equation}}
\newcommand{\bea}{\begin{eqnarray}}
\newcommand{\eea}{\end{eqnarray}}

\newcommand{\CMP}{{\it Commun. Math. Phys.\,}}
\newcommand{\CQG}{{\it Class. Quantum Grav.\,}}
\newcommand{\IJMP}{{\it Int. J. Mod. Phys.\,}}

\newcommand{\JMP}{{\it J. Math. Phys.\,}}

\newcommand{\NP}{{\it Nucl. Phys.\,}}
\newcommand{\PL}{{\it Phys. Lett.\,}}
\newcommand{\PR}{{\it Phys. Rev.\,}}

\newcommand{\ZP}{{\it Z. Phys.\,}}

\newcommand{\T}{\text{O}}
\newcommand{\F}{\text{X}}
\newcommand{\hcite}[1]{$\text{ }$\cite{#1}}

\begin{document}

\title{Quantization of Spacetime Based on Spacetime Interval Operator}

\author{Hsu-Wen~Chiang}
\email{b98202036@ntu.edu.tw}
\affiliation{Department~of~Physics, National~Taiwan~University, Taipei~10617, Taiwan, R.O.C.\\}
%\affiliation{Graduate~Institute~of~Physics, National~Taiwan~University, Taipei~10617, Taiwan, R.O.C.\\}
\affiliation{Leung~Center~for~Cosmology~and~Particle~Astrophysics, National~Taiwan~University, Taipei~10617, Taiwan, R.O.C.\\}

\author{Yao-Chieh Hu}
\email{r04244003@ntu.edu.tw}
\affiliation{Department~of~Physics, National~Taiwan~University, Taipei~10617, Taiwan, R.O.C.\\}
\affiliation{Leung~Center~for~Cosmology~and~Particle~Astrophysics, National~Taiwan~University, Taipei~10617, Taiwan, R.O.C.\\}
\affiliation{Graduate~Institute~of~Astrophysics, National~Taiwan~University, Taipei~10617, Taiwan, R.O.C.\\}

\author{Pisin Chen}
\email{pisinchen@phys.ntu.edu.tw}
\affiliation{Department~of~Physics, National~Taiwan~University, Taipei~10617, Taiwan, R.O.C.\\}
%\affiliation{Graduate~Institute~of~Physics, National~Taiwan~University, Taipei~10617, Taiwan, R.O.C.\\}
\affiliation{Leung~Center~for~Cosmology~and~Particle~Astrophysics, National~Taiwan~University, Taipei~10617, Taiwan, R.O.C.\\}
\affiliation{Graduate~Institute~of~Astrophysics, National~Taiwan~University, Taipei~10617, Taiwan, R.O.C.\\}
\affiliation{Kavli~Institute~for~Particle~Astrophysics~and~Cosmology, SLAC~National~Accelerator~Laboratory, Stanford~University, Stanford, CA~94305, U.S.A.\\}

\begin{abstract}
Motivated by both concepts of R.J. Adler's recent work on utilizing Clifford algebra as the linear line element $ds = \left\langle \gamma _\mu \right\rangle dX^\mu $, and the fermionization of the cylindrical worldsheet Polyakov action, we introduce a new type of spacetime quantization that is fully covariant. The theory is based on the reinterpretation of Adler's linear line element as $ds = \gamma _\mu \left\langle \lambda \gamma ^\mu \right\rangle$, where $\lambda$ is the characteristic length of the theory. We name this new operator as ``spacetime interval operator", and argue that it can be regarded as a natural extension to the one-forms in the $U(\mathfrak{s}u(2))$ non-commutative geometry. By treating Fourier momentum as the particle momentum, the generalized uncertainty principle of the $U(\mathfrak{s}u(2))$ non-commutative geometry, as an approximation to the generalized uncertainty principle of our theory, is derived, and is shown to have a lowest order correction term of the order $p^2$ similar to that of Snyder's. The holography nature of the theory is demonstrated, and the predicted fuzziness of the geodesic is shown to be much smaller than conceivable astrophysical bounds.
\end{abstract}

\maketitle
\section{Introduction}
\label{sec:intro}
In Riemannian geometry, the quadratic distance function $g$ living on a manifold $M$ describes the structure of the tangent bundle $TM$ uniquely through specifying the relation between quadratic line element $ds_R^2$ and the coordinate difference $dx$. The quadratic nature of $ds_R^2 = g_{\mu\nu} dx^\mu dx^\nu$ which implies Lorentz symmetry and Pythagorean theorem, is based on numerous experimental facts\hcite{LorentzInvExp}. It is one of the foundations of general relativity (GR) and even when quantizing gravity, people usually promote it to its quantum version without modification. But in both loop quantum gravity (LQG) and superstring theories, the existence of a minimal distance measure suggests that the infinitely-differentiable geometry may be an illusion that ceases to be valid at the smallest scale. This interesting consequence of combining GR with quantum field theory (QFT) leads to the notion that spacetime structure itself may have to be modified. H.S. Snyder coined this attempt in the name of ``quantized space-time" in his seminal article published in 1947\hcite{Snyder}.

Currently there are two major routes to tackle the quantization of spacetime. Dimensional reduction from a higher dimensional momentum space proposed by Snyder is a popular approach, which was followed by S.~Majid, G.~Amelino-Camelia, and others in the construction of their own versions of quantized spacetime\hcite{Bicrossproduct,DSRIntro,DSRBasis}. The other route, first introduced by A.~Connes\hcite{Connes1,Connes2}, comes from partial differential equation (PDE) analysis on non-commutative space ($C^*$ algebra), where fields and Fourier analysis can be defined classically, with twisted measure providing non-commutativity. This approach has been proven to be extremely versatile in the pursuit of quantized spacetime as non-commutativity usually plays an important role\hcite{SU2F,SL2CF,3DQG,ConstNoncomm}.

Here we consider a new way of deforming spacetime algebra, first proposed by R. Adler\hcite{Adler1}, which has its root in Clifford algebra of the tangent bundle. Instead of a bosonic structure where generators of momentum space serve as the coordinate measures, this theory treats the proper distance measure as the composition of infinite generators of Clifford structure on the tangent bundle of the position space. A continuous geodesic therefore becomes piecewise-linear and so is the manifold. As we will show by carefully defining the measure, a quantum mechanical (QM) system can be constructed on top of it.

This paper is structured as following. In sec.\ref{sec:review}, we will review some important quantized spacetime models that are related to our work. In sec.\ref{sec:foundation}, the fundamental structure of the theory is presented. In sec.\ref{sec:action}, we utilize the bosonization technique to relate our theory to the conventional relativistic QM. In sec.\ref{sec:commutator}, the position operator is constructed. The scaling law for the number of states within a certain volume and the associated holographic principle are derived as well. In sec.\ref{sec:GUP}, a form of approximate momentum operator, Heisenberg commutation relation and the generalized uncertainty principle (GUP) are presented. In sec.\ref{sec:astrophystest}, the smearing effect of our theory is deduced, and finally the conclusion is made in sec.\ref{sec:conclusion}. For the sake of readability, when using subscripts/superscripts, unless defined specifically, $IJKL$ running from 0 to 3 are for $SO(3,1)$ tangent bundle coordinate indices, $\mu \nu \lambda \rho$ from 0 to 3 for 4-$d$ coordinate indices, $i j k l$ from 1 to 3 for 3-$d$ coordinate indices, $a b c d$ from 1 to 3 for 3-$d$ tangent bundle indices, $\alpha \beta \gamma \delta$ from 0 to 1 for 2-$d$ coordinate indices, and $p q r s$ from 0 to 3 for 4-$d$ spin indices. We also use natural units $c = \hslash = G = 1$ to simplify equations, and the signature of metric $g$ is chosen to be $\left(+,-,-,-\right)$.

\section{Review on Quantization of Spacetime}
\label{sec:review}
Snyder suggested\hcite{Snyder} a deformation stemmed from the exponential map of a 4-$d$ de Sitter ($dS$) space with "radius" $a^{-1}$ embedded inside a 5-$d$ momentum space (~$dk_{\mathbb{R}^{1,4}}^2 = dk_0^2 - dk_1^2 - dk_2^2 - dk_3^2 - dk_4^2$~),~i.e.,
\bea
a^{-2} = k_0^2 - k_1^2 - k_2^2 - k_3^2 - k_4^2 \;.
\eea
Next, $\hat{x}^\mu$ is chosen to be the momentum translation Killing vector such that it satisfies Lorentz symmetry. Then the deliberately chosen conformally flat hypersurface guarantees that the Lie derivatives of the Killing vectors must be proportional to the Lorentz transformation Killing vectors $J^{\mu\nu}$:
\bea
\hat{x}^\mu &=& i a \left( k_4 \frac{\partial}{\partial k_\mu} + k^\mu \frac{\partial}{\partial k_4} \right) \,, \\
\left[ \hat{x}^\mu , \hat{x}^\nu \right] &=& i a^2 \hat{J}^{\mu\nu} = - a^2 \left( k^\mu \frac{\partial}{\partial k_\nu} - k^\nu \frac{\partial}{\partial k_\mu} \right) \,,\ \ 
\eea
where $k^\mu = \eta ^{\mu\nu} k_\nu$, $\eta ^{\mu\nu}$ is the 4-$d$ Minkowski metric and $\left[ \;\, , \; \right]$ is the commutator. The momentum coordinate is the exponential map,
\bea
\hat{p}_\mu &=& a^{-1} k_\mu / k_4 \;,
\eea
on which the Lorentz transformation Killing vectors $\hat{J}^{\mu\nu}$ locally have the same form as the traditional Lorentz transformation generators $\hat{J}^{\mu\nu} = \hat{x}^\mu \hat{p}^\nu - \hat{x}^\nu \hat{p}^\mu$. Snyder's approach therefore can be viewed as a non-abelian realization of the Lorentz group. The Heisenberg relation and the uncertainty relation are twisted accordingly to be
\bea
\left[ \hat{x}^\mu , \hat{p}_\nu \right] &=& i \left( \delta ^\mu _\nu - a^2 \eta ^{\mu\lambda} \hat{p}_\lambda \hat{p}_\nu \right) \,,\\
\Delta x^\mu \Delta p_\nu &\geq & \frac{1}{2} \left( \delta ^\mu _\nu - a^2 \eta ^{\mu\lambda} \left\langle \hat{p}_\lambda \hat{p}_\nu \right\rangle \right) \,. \label{snyderGUP}
\eea
As we will see in sec.\ref{sec:GUP}, our own version of quantized spacetime carries the same uncertainty relation as Snyder's up to the first order in $p^2$.

An weaker deformation, called $\kappa$-Poincar\' e group, later invoked by G. Amelino-Camelia\hcite{DSRIntro,DSRBasis} for the doubly special relativity, was first introduced by S. Majid and H. Ruegg\hcite{Bicrossproduct} as follows:
\bea
\left[ x^i , x^j \right] = 0 \,,\quad \left[ x^i , x^0 \right] = \kappa ^{-1} x^i \,,\quad \left[ p_\mu , p_\nu \right] = 0 \,.\,
\eea
This system can be regarded as an $AdS_3 \times \mathbb{R}^{1,0}$ hypersurface in a 5-$d$ momentum space, where an upper bound $\kappa$ for the spatial momentum $\left| \vec{p} \right|$ served as a cutoff scale for the QFT built on top of it. The realization of a maximally symmetric momentum space and the flat $p_0$ direction implies a vanilla Lorentz group constructed from $\hat{p} \hat{x}$ and a linear action on $p_0$. Note, however, that the cylindrical structure implies a non-linear Lorentz boost $N_i$ on the spatial momentum
\bea
\left[ N_i , p_j \right] = i \delta _{ij} \left[ \frac{\kappa}{2} \left( 1 - e^{-2 p_0 / \kappa} \right) + {\left| \frac{\vec{p}}{2} \right|}^2 \right] - \frac{i}{\kappa} p_i p_j \, . %{\color{white}.}
\eea
A maximally achievable spatial momentum $\kappa$ therefore corresponds to an infinite energy $p_0$. In sec.\ref{sec:GUP} , we will demonstrate that there also exists a maximal spatial momentum with finite energy in our theory. However, their commutation relation and accordingly their generalized uncertainty relation, share no similarity to our result.

Another work that is closely related to what we propose here is the angular momentum space theory introduced by Shahn Majid\hcite{SU2F}, where the group structure of the angular momentum operator, $SU(2)$, is considered as a deformation of the 3-$d$ spacetime
\bea
\left[ x^i , x^j \right] = 2 i \lambda \epsilon ^{ij}_k x^k \,. \label{su2xcommute}
\eea
The most surprising feature of this theory is that the construction of the differential structure through $C^*$ algebra implies the existence of a fourth direction, which supposedly is the time. Assuming the completeness of the Lie algebra, the commutation relation of the 1-form and the exterior derivative of the Fourier mode are
\bea
\left[ x_i , dx_j \right] = i \lambda \epsilon _{ijk} dx_k + \lambda \delta _{ij} d\tau \,,\;  \left[ x_i , d\tau \right] = \lambda dx_i &\ \ & \label{su2xdxcommute} \\
d e^{i k\cdot x} = \left[ \frac{i \sin \left( \lambda k \right) }{\lambda k} k \cdot dx - \frac{8 d\tau}{\lambda} \sin ^2 \left( \frac{\lambda k}{2} \right) \right] e^{i k\cdot x} &\ \ & \label{su2dexp}
\eea
where $d$ stands for the exterior derivative, $dx_i = \sigma _i /2$ and $d\tau = I_2/2$, $\sigma _i$ are the Pauli matrices, $I_2$ is the identity operator, $k$ is the Fourier momentum with $k^2=k_a k^a$, and $k\cdot x=k_a x^a$. Majid regarded this theory as a 3-$d$ remnant of $q$-deformed poincar\' e group theory\hcite{qdeform}. We will show that what we propose is another way of extending the theory to 4-$d$, by regarding $SU(2)$ group as Clifford algebra on 3-$d$ Euclidean manifold, or by focusing on connected subgroup of the $Spin(3,1)$.

Following a completely different reasoning, R. Adler arrived at a quantized spacetime as follows. Considering the Riemannian distance functional $ds_R^2 = g_{\mu\nu} dx^\mu dx^\nu$, a natural way to get the operator spectrum, other than the spectral square root\hcite{LQGlength0}, would be via Clifford algebra. One then obtains a linear functional $ds_L = \gamma _\mu dx^\mu$ connecting the coordinate difference to the linear line element. The prize to pay is that the new transfer functional $\gamma _\mu$, which was chosen to be the chiral gamma matrix, is a matrix instead of a pure number.

To compensate that, Adler reinterprets the distance functional as a QM object, called ``linear line element operator", whose eigenvalue is the proper distance and the state describes an eigen-direction along which the performed measure is completely certain.
\bea
ds_A = \langle \hat{ds} \rangle &=& \left\langle \epsilon \left| \gamma _\mu dx^\mu \right| \epsilon \right\rangle \label{ds_A} = \epsilon _\mu dx^\mu \,,\\
\left\langle \epsilon \left| \right| \epsilon \right\rangle &=& \left( \left| \epsilon \right\rangle \right) ^\dagger  \left| \epsilon \right\rangle = 1 \;, \label{ds_A:measure}
\eea
where $\hat{ds}$ stands for the linear line element operator, $\gamma _\mu$ are the chiral gamma matrices, and $\epsilon _\mu$ is the expectation value of the distance functional. The subscript $A$ in $ds_A$ indicates that this is Adler's version of line element. In addition, Adler assumes that the traditional picture of the infinitely differentiable manifold should be replaced by a piecewise-linear manifold with the curve length
\bea
L_C = \sum _i \big{\langle} \hat{\Delta s} \big{\rangle} _{\left( i \right)} = \sum _i \left\langle \epsilon _{\left( i \right)} \right| \gamma _\mu \Delta x^\mu_{\left( i \right)} \left| \epsilon _{\left( i \right)} \right\rangle \,,
\eea
where $\left( i \right)$ indicates the site and $\Delta x = \lambda dx$ is the discrete coordinate difference. So instead of the Killing vector of a curved momentum space serving as the coordinate measure or a twisted commutation relation of position operator, the proper distance is treated as the addition of generators of the spin group on classical x-space, similar to the 't HOOFT operator on a spacetime lattice.

This theory introduces an uncertainty to the proper distance measure as
\bea
\Delta L_C^2 = \sum _i  \left[ 1- { \left( \dot{x}^{\mu\left( i \right)} \epsilon _{\mu\left( i \right)} \right) }^2 \right] g_{\mu\nu} \Delta x^\mu_{\left( i \right)} \Delta x^\nu_{\left( i \right)} \,,
\eea
where $\dot{x}^\mu = dx^\mu / ds$ is the $4$-velocity, and $\epsilon _\mu$ is the preferred direction that minimizes the uncertainty. The notion of a preferred direction inevitably breaks the Lorentz invariance, unless the theory is non-dissipative or is at high temperature limit. As such, the whole purpose of using Clifford algebra to preserve Lorentz invariance is in vain. The failure of Adler's original attempt prompts us to construct a different realization of the same idea.

\section{Spin Structure as Spacetime Quantum Crystal and Solution to the Arrow of Time}
\label{sec:foundation}
Following Adler's argument, the Riemannian line element $ds_R^2 = g_{\mu\nu} dx^\mu dx^\nu$ can be discretized and factorized via the measurement of the square of the linear line element 
\bea
\Delta s_A^2 = \big{\langle} \hat{\Delta s^2} \big{\rangle} = \left\langle \epsilon \left| \gamma _\mu \Delta x^\mu \gamma _\nu \Delta x^\nu \right| \epsilon \right\rangle \,,
\eea
with
\bea
{ \left\{ \gamma ^I , \gamma ^J \right\} }_{pq} &=& 2 \eta ^{IJ} \otimes I_{pq} \,,\\
{ \left[ \gamma ^I , \gamma ^J \right] }_{pq} &=& -2 i \sigma ^{IJ}_{pq}\,,\label{sigma} \\
\gamma ^\mu &=& e_I^\mu \gamma ^I \,,\\
g_{\mu\nu} = \eta _{IJ} e^I_\mu e^J_\nu &,& \eta _{IJ} = g^{\mu\nu} e^I_\mu e^J_\nu \,,
\eea
where $e$ is the tetrad field and $\left\{ \;\, , \; \right\}$ is the anti-commutator.

Care must be taken when one deals with the definition and the interpretation of the measure. In Adler's approach, Eq.(\ref{ds_A:measure}) was used to define the braket, while Eq.(\ref{ds_A}) was interpreted as a distance functional. However, this choice suffers some drawbacks and is unsuitable for the construction of our quantum spacetime theory.

First, the linear line element operator contains the exact information one would expect to be hidden inside the Hilbert space of the quantized spacetime, i.e., the direction. In the original interpretation, the Hilbert space contains the information for the uncertainty of distance measurement rather than the direction itself. The direction of the line element is provided externally in Eq.(\ref{ds_A}) since the theory is describing a quantized distance functional, not a quantized spacetime. The existence of a favored direction that minimizes the uncertainty also breaks Lorentz invariance and isotropy at the smallest scale. The salient feature of Lorentz symmetry in Adler's theory (Dirac's way of taking square root clearly is Lorentz invariant) is therefore lost.

Second, when one measures the proper distance of null eigenstates (which should be quite common given the fact that all particles are massless prior to electro-weak symmetry breaking) along any direction, because of the choice of normalization the proper distance would always be zero. So for a null state even if the curve is not along null direction the proper distance would still be null. To wit, the measure of proper distance is completely uncertain for a non-null displacement on a null state.

Third, the outcome of the linear distance functional depends heavily on the choice of the representation of the Clifford algebra. A complex representation could result in a complex proper distance, which would be a radical departure from GR. Although Adler tries to address this issue by fixing the representation, the problem still exists as long as the proper distance, being a physical measure, is not a scalar of Clifford algebra, i.e. not representation independent.

Clearly, these observations indicate the necessity of reinterpretation of Adler's linear line element and a new choice of normalization. We look for a new definition that should satisfy Lorentz symmetry, should not have preferred direction, should produce reasonable results for null states, and should be independent of the choice of the representation.

\begin{table*}
\begin{align}
\begin{array}{c|cccccccc}
 &\Delta \vec{X}&\Delta V_3&\Delta \vec{V}&\Delta t&\Delta s^2&\Delta \vec{A}&\Delta \vec{A_t}&\Delta V_4\\ \hline
\Delta \vec{X}  & \T & \T & \T & \T & \F & \F & \F & \F \\
\Delta V_3      & \T & \T & \T & \T & \F & \F & \F & \F \\
\Delta \vec{V}  & \T & \T & \T & \T & \T & \T & \T & \T \\
\Delta t        & \T & \T & \T & \T & \T & \T & \T & \T \\
\Delta s^2      & \F & \F & \T & \T & \T & \T & \F & \F \\
\Delta \vec{A}  & \F & \F & \T & \T & \T & \T & \F & \F \\
\Delta \vec{A_t}& \F & \F & \T & \T & \F & \F & \T & \T \\
\Delta V_4      & \F & \F & \T & \T & \F & \F & \T & \T \\
\end{array} \nonumber
\end{align}

\caption{A commutativity table showing possible ways of labelling Hilbert space. For elements $T_{mn}$ inside the table, ``$\T$" means $m$-th basis commutes with $n$-th basis, and ``$\F$" means non-commutativity. Here all vectors are along spatial eigen-direction $n^i = \left\langle \Delta X^i \right\rangle$, and $\Delta \vec{X} = n_i \Delta X^i$ is the spatial interval, $\Delta V_3 = \Delta X^1 \Delta X^2 \Delta X^3$ is the time-like 3-volume, $\Delta \vec{V} = - n_i \epsilon ^i_{jk} \Delta X^0 \Delta X^j \Delta X^k$ is the spatial 3-volume, $\Delta t$ is the time difference, $\Delta s^2$ is the proper distance square, $\Delta \vec{A} = n_i \epsilon ^i_{jk} \Delta X^j \Delta X^k$ is the spatial area, $\Delta \vec{A_t} = n_i \Delta X^0 \Delta X^i$ is the time-like area, and $\Delta V_4 = \Delta X^0 \Delta X^1 \Delta X^2 \Delta X^3$ is the 4-volume. Notice that actually $\Delta \vec{V}$ can always be described by products of two non-trivial quantum numbers in the system.}
\label{BasisCommuteTable}
\end{table*}

So we give up Eq. (\ref{ds_A:measure}) and introduce a new operator $\Delta \hat{X}^I$, called ``spacetime interval operator":
\begin{align}
\Delta \hat{X}^I &= e^I_\mu \Delta \hat{X}^\mu = \lambda \gamma ^I \,,\\
\big{\langle} \Delta \hat{X}^I \big{\rangle} &= \bar{\psi} \Delta \hat{X}^I \psi = \psi ^{\dagger} \gamma ^0 \Delta \hat{X}^I \psi \,, \label{measure} \\
\bar{\psi} \psi &= \left\{\begin{array}{ll}
1   &     ,\; \forall \text{ time-like states}\\
0   &     ,\; \forall \text{ null states}\\
-1  &     ,\; \forall \text{ space-like states}\\
\end{array}\right. \label{normalization} \\
\bar{\psi}\gamma ^I \psi &= \left\{\begin{array}{ll}
n^I & \,\,,\; \forall \text{ non-null states}\\
k^I & \,\,,\; \forall \text{ null states}\\
\end{array}\right. \label{X:expectation}\\
ds^2 &= \eta _{IJ} \left\langle \Delta \hat{X}^I \Delta \hat{X}^J \right\rangle = 4\lambda ^2 \bar{\psi} \psi\,.
\end{align}
Here $\lambda$ is the characteristic length of the quantized spacetime that is of the order of Planck length and will be derived in Sec.\ref{sec:action}, $n^I$ is a non-null vector with $n_I n^I = \pm 1$, and $k^I$ is a null vector with positive $k^0$. The appearance of the $\gamma ^0$, and the choice of normalization can be appreciated by looking at the solution of Dirac field equation (See, for example, Ch.~3-3 of Peskin~\&~Schroeder\hcite{Peskin}.) Physically the insertion of $\gamma ^0$ makes the normalization condition Lorentz invariant, and for the massless case the choice of normalization is equivalent to the introduction of the foliation along the time coordinate.

This new measurement is not a distance measure at all, but a local spacetime interval operation on the exponential map of Riemann Normal Coordinate. One should not misinterpret the operator as a coordinate difference operator, since it actually lies on the tangent bundle of the manifold. A better way to understand it is to treat it as a discretized version of the velocity 4-vector. Only when combined with the tetrad does the coordinate difference measure $\Delta \hat{X}^\mu$ reappear.

Just like the components of a vector in GR, $\Delta \hat{X}^I$ (gamma matrices) are not physical objects. To measure the velocity of a particle one needs a two-particle interaction, and the physical object is the inner product $r_I \Delta \hat{X}^I$, where $r_I$ is the classical trajectory of a probe particle expressed in the same representation as $\Delta \hat{X}^I$. No matter what representation of Clifford algebra that we choose, the measurement is always a scalar. Therefore all the derivations and results we obtained are representation independent. The only exception is that in Sec. VI, where for convenience we require the realness of $\Delta \hat{X}^I$ in our derivation of GUP. However, one should obtain the same GUP regardless of the representation used.

The new choice of interpretation also implies the existence of an underlying minimal distance. Due to the special structure of Clifford algebra, one can immediately obtain $\left| \left\langle \Delta X^I \Delta X^I \right\rangle \right| = \lambda ^2$ for non-null cases and $0$ for null cases, implying that there are uncertainties within the spacetime interval measurement similar to what was obtained in Ref.\cite{Adler1}, where the uncertainty lies on proper distance measurement. But in our case even such uncertainties are Lorentz invariant. One may try to obtain the variance of interval measure:
\bea
\left\langle \text{var} \left( r_I \Delta X^I \right) \right\rangle &=& \left| \left\langle {\left( r_I \Delta X^I \right) }^2 \right\rangle -{\left\langle r_I \Delta X^I \right\rangle }^2 \right| \nonumber \\
&=& \lambda ^2 \left| r \cdot r \times n \cdot n - {\left( r \cdot n \right) }^2 \right| \,, \label{var_dX} \\
\left\langle \text{var} \left( ds_R \right) \right\rangle &=& \left| \left\langle \Delta X^I \Delta X_I \right\rangle - \left\langle \Delta X^I \right\rangle \left\langle \Delta X_I \right\rangle \right| \nonumber \\
&=& \left\{ \begin{array}{ll}
3 \lambda ^2 & ,\; \forall \text{ non-null states} \\
0            & ,\; \forall \text{ null states} \\
\end{array} \right. \label{var_ds}
\eea

Here $n^I = \left\langle \Delta X^I \right\rangle$, $r^I$ is the ruler, and $A \cdot B = \eta _{IJ} A^I B^J$ is the inner product. Clearly along eigen-direction there is no uncertainty at all since it is the direction where eigenstates are defined. However along the transverse direction the measurement is completely uncertain. From this point of view the behaviour of the spacetime interval operator is exactly the same as spin operators in relativistic QM. They both have 3 definite quantum numbers, i.e., $S^2$/$\Delta t$, $\vec{S}$/$\Delta \vec{X}$ along spatial eigen-direction, and helicity/$\Delta V_3$ the spatial 3-volume. There are also other possible ways of labelling Hilbert space, which are shown in Table \ref{BasisCommuteTable}.

N\"{a}ively, in $SO(3,1)$ system there are 4 quantum numbers: the 4-momentum. However they do not correspond to the quantum numbers in our theory since the momentum operators do not commute with each other. Only at the decoherence limit will the additional 4 quantities, eigen-direction, spatial 3-volume along eigen-direction and proper distance, emerge.

An important feature of Adler's original interpretation is that the proper time difference $\Delta s \propto \gamma ^\mu$ can have two eigenvalues of same magnitude but opposite sign. Therefore in his theory it is permissible to move backward in time, zigzag around the same spot at high temperature if no rule forbids the excitation of these states. However in our interpretation, time difference is now proportional to the identity operator due to the choice of measure, rendering its expectation value positive definite. Thus arrow of time problem is perfectly solved without invoking second law of thermodynamics.

In Ref.\cite{Adler1}, Adler suggests that one may take these tiny line elements as building blocks of a curve, without specifying what kind of curve it is. Since in the original paper the measurement of linear line element operators are associated with $SO(3,1)$ proper distance, this curve resembles a geodesic, called ``quantized geodesic" due to its discrete nature. Under our spacetime interval operators, the link to the geodesic becomes more explicit since a trajectory is specified along the geodesic as

\bea
\left| \mathcal{C} \right\rangle = \underset{i}{\oplus} \left| n_{ \left( i \right) } \right\rangle &,& \; n^I_{ \left( i \right) } = \left\langle n_{ \left( i \right) } \right| \Delta X^I \left| n_{ \left( i \right) } \right\rangle \, ,\\
X^\mu _{ \left( i \right) } = \sum _{j=1}^i e^\mu _{ I \left( j \right) } n^I_{ \left( i \right) } &,& \;
s = \sum _i \sqrt{ n^I_{ \left( i \right) } n_{ I \left( i\right) } } \, .
\eea
Here $\left| \mathcal{C} \right\rangle$ is the composite state for geodesic, and $\left( i \right)$ indicates the $i$-th site along the geodesic. Following the usual convention in special relativity, the positive eigenvalue of proper distance $\Delta s = \pm \sqrt{\Delta s^2}$ is chosen due to the positivity of the time difference.

%It is straightforward to apply quantized geodesic to the trajectory of cosmic ray, which may serve as astrophysical test for this theory through smearing of particle trajectory. Looking at eq.(\ref{var_dX}) and eq.(\ref{var_ds}), for null geodesic, eg. GRB photon, since variance occurs only in perpendicular direction, assuming Gaussian distribution, smearing of the photon is of order $\sqrt{ \lambda \; d_C \left( z<1100 \right) } < 0.6$ light hour, where $\lambda$ is the GRB characteristic wavelength which must be smaller than background CMB photon, and $d_C$ is the comoving distance. Clearly it's impossible to identity such tiny effect in near future.

%For time-like geodesic, in addition to perpendicular direction, there are also variance in proper distance, which allow us for more precise measurement. The smearing is at most $\sqrt{3 t_P \; T \left( z=1100 \right) } \approx 1$ nanosecond, where $t_P$ is the Planck time and $T$ is the time taken from $z=1000$ to $z=0$. Therefore it is almost impossible to measure this effect unless events like primordial blackhole evaporation occur.

\section{Constructing Action through Bosonization: A Low Energy Limit of Worldsheet Theory}
\label{sec:action}
Up until now our theory is still presented in a schematic way where the operators and the measure were not derived from first principles. Given Eq.(\ref{measure}), we can deduce that a worldline action in terms of bi-fermionic fields exhibiting bosonic behavior should describe our theory. A natural tool to attain this action would be the bosonization in 2-$d$ QFT\hcite{bosonization}. By considering the worldline as the low energy limit of a cylindrical worldsheet, we may apply the fermionization on the worldsheet action and compactify it back to 1-$d$.

Considering the worldline action for a relativistic particle
\bea
S_R &=& \int \left( \xi g_{\mu \nu} \frac{dX^\mu }{d \tau} \frac{dX^\nu }{d \tau} - \frac{ m^2}{ 2 \xi} \right) \, d \tau \,,
\eea
where $\tau$ is a parameter along the particle trajectory, and $\xi$ is an axillary field with equation of motion
\bea
\xi &=& m \left( g_{\mu \nu} \frac{dX^\mu }{d \tau} \frac{dX^\nu }{d \tau} \right)^{-1/2} \,.
\eea
One can describe it as the Kaluza-Klein ground state of the worldsheet Polyakov action on a cylinder through compactification along small perimeter $L \ll m$
\bea
S_R &\approx & \int _0^L d \sigma ^1 \frac{ S_R + c(\sigma ^1)}{L} \approx S_P \,,\\
S_P &=& \frac{1}{8 \pi l_s^2} \int d^2 \sigma \sqrt{-h} h^{\alpha \beta} g_{\mu \nu} \partial _\alpha X^\mu \partial _\beta X^\nu \,,
\eea
where $l_s = \sqrt{L / \left(4 \pi m \right)}$ is the string length, $\sigma ^0 = \tau$, and $\sigma ^1$ is the compactified direction. Up to now we still cannot apply the fermionization. A na\" \i vely fermionized action would be ambiguous as the group structure of the system $g\left(X\right)$ is still undetermined. To avoid such problem one may rely on tetrad formalism to pull $g$ back to Minkowski metric $\eta$ and obtain a $SO \left( 3,1 \right)$ theory
\begin{align}
S_P &= \frac{1}{8 \pi l_s^2} \int d^2 \sigma \sqrt{-h} h^{\alpha \beta} \nabla ^I_{K \alpha} X^K \nabla _{IL \beta} X^L \,, \\
X^I &= e^I_\mu X^\mu \,,\; \nabla _\alpha = \partial _\alpha + \omega ^I_{J\alpha} - \Gamma ^I_{J\alpha} \,,
\end{align}
where the tetrad field $e$ should be a functional of $X$, $\nabla _\alpha$ is the covariant derivative which annihilates the tetrad $\nabla _\alpha e = 0$, $\omega ^I_{J \alpha}$ is the antisymmetric spin connection, and $\Gamma ^I_{J \alpha}$ is the symmetric affine connection. The $SO \left( 3,1 \right)$ spacetime group can now be replaced with the $Spin\left(3,1\right)$ group, making this theory a spacetime bi-fermion theory. The new action under the gauge condition $\Gamma ^I_{I\alpha} = 0$ would be
\begin{align}
S_{Spin} &= \frac{1}{8 \pi} \int d^2 \sigma \sqrt{-h} D _\alpha \slashed{\chi} D ^\alpha \slashed{\chi} \,, \label{S_Spin} \\
D_\alpha \slashed{\chi} & = l_s^{-1} \gamma _I e^I_\mu \partial _\alpha X^\mu \,, \;
         \slashed{\chi}   = l_s^{-1} \gamma _I e^I_\mu                  X^\mu \,, \\
D_\alpha & = \partial _\alpha + \frac{1}{2} \omega ^I_{J\alpha} \left[ \gamma _I , \gamma ^J \right] = \partial _\alpha - i \omega ^I_{J\alpha} \sigma _I^J \,,
\end{align}
where $\slashed{A} = \gamma _I A^I$ and $D_\alpha \slashed{\chi}$ is the generalized velocity of the $Spin \left( 3,1 \right)$-valued worldsheet scalar $\slashed{\chi}$ of dimension $[m^0]$. Notice that $D_0 \slashed{\chi}$ is exactly the continuum limit of Adler's linear line element operator, indicating that we are on the right track. Since the mapping from worldsheet space to $Spin \left( 3,1 \right)$ space is trivial for K-K ground state, both E. Witten's approach\hcite{nonabelianbosonization} and Steinhardt's\hcite{Steinhardt} preserve the symmetry after fermionization. Therefore one can fermionize the theory by checking the bozonization dictionary:
\bea
4\pi \bar{\eta} \left( 1 \pm \sigma ^2 \right) \eta  &\Longleftrightarrow & e^{\pm 4 \pi i \phi} \label{dict} \\
\bar{\eta} \sigma ^\alpha \eta &\Longleftrightarrow & -2 \epsilon ^{\alpha \beta} D_\beta \phi \\
i \bar{\eta} \sigma ^\alpha D_\alpha \eta  &\Longleftrightarrow & D_\alpha \phi D^\alpha \phi 
\eea
, where $\phi$ is a real scalar of dimension $[m^0]$, $\eta$ is a 2-$d$ worldsheet Majorana spinor of dimension $[m^0]$, $\sigma ^\alpha$ are the 1+1-$d$ Majorana-Weyl matrices , and $\sigma ^2 = \sigma ^0 \sigma ^1$. For the sake of simplicity we have rescaled both $\eta$ and $\phi$ by a factor of $\sqrt{4\pi}$. Applying these rules to Eq.(\ref{S_Spin}) one arrives at
\bea
S_F &=& \int d^2 \sigma \sqrt{-h} \left( i \bar{c} \sigma ^\alpha D_\alpha c \right) \,,
\eea
where $c$ is a $4$-d spacetime spinor valued $2$-d worldsheet Majorana-Weyl spinor.

From constraints of the K-K ground state and the bosonization dictionary one immediately obtains
\bea
D_1 \slashed{\chi} &=& 0           = - \frac{1}{2} c^\dagger           c \,,\\
D_0 \slashed{\chi} &=& \slashed{v} =   \frac{1}{2} c^\dagger \sigma ^2 c \,, \label{constraint}
\eea
where $\slashed{v}$ is the velocity.
To satisfy these two requirements the only possible solution would be
\bea
c = \left( \psi , \psi C \right) = \left( \sqrt{\gamma ^0 \slashed{v}} \psi _0 , \sqrt{\gamma ^0 \slashed{v}} \psi _0 C \right) \,,
\eea
where $\psi _0$ is a normalized spinor with $\psi _0^\dagger \psi _0 = 1$, and $C$ is the usual hermitian charge conjugation operator that transforms $\bar{\psi} \gamma ^\mu \psi$ to $-\bar{\psi} \gamma ^\mu \psi$. Finally we can obtain the compactified action and the geodesic equation in terms of fermionic field $\psi$
\bea
S_C &=& \int d \tau \left( i \bar{\psi} D_0 \psi \right) \,,\\
\frac{d^2 X^\mu}{d \tau ^2} &=& \frac{d}{d \tau} \bar{\psi} e^\mu _I \gamma ^I \psi = - e^\mu _I \Gamma ^I_{J0} v^J \,. \label{eom}
\eea
Since the worldsheet spinor $\bar{c} c$ is conserved, from Eq.(\ref{dict}) and the identity
\bea
e^{i \slashed{F} } = \cos \left( \sqrt{F_I F^I} \right) I_4 + i \sin \left( \sqrt{F_I F^I} \right) \slashed{F} \,, \label{expidentity}
\eea
we obtain
\bea
\slashed{\chi} = \left( \frac{n}{2 \sqrt{\left| v^2 \right|}} + q \right) \slashed{v} \quad ,\; n \in \mathbb{Z} \,,
\eea
where $v^2 = v_I v^I$, $q$ is a scalar constant in terms of $v^2$, and spacetime is quantized accordingly with the characteristic length
\bea
\lambda = \frac{1}{2} \, l_s \,.
\eea
%The additional $\pi$ will be shown to be important in Sec.\ref{sec:holographic} when we fix the Bekenstein-Hawking coefficient.
Eqs.(\ref{measure}) to (\ref{X:expectation}) can be re-derived respectively as
\begin{align}
\big{\langle} \Delta \hat{X}^\mu \big{\rangle} &= \lambda \bar{\psi} \gamma ^\mu \psi = \frac{\lambda}{4} tr \left( \gamma ^\mu \slashed{v} \right) = \lambda v^\mu \,,\\
v^2 = \frac{{\left| \big{\langle} \Delta \hat{X}^\mu \big{\rangle} \right|}^2}{\lambda ^2} &\approx 4 \Delta \slashed{\chi} \Delta \slashed{\chi} = \frac{v^2}{\left| v^2 \right|} = \text{sign} \left( v^2 \right) \,,\\
\bar{\psi} \psi &= \frac{\lambda}{4} tr \left( \slashed{v} \right) = 0 \,.
\end{align}
Since in this section we focus on eigenstates of the spacetime interval operator $\Delta \hat{X}^I$, $\bar{\psi} \psi$ must vanish. If the constraint Eq.(\ref{constraint}) is substituted with $\bar{\psi} \psi = const.$, from Eq.(\ref{dict}) and (\ref{expidentity}) we immediately obtain
\bea
\bar{\psi} \psi = 1 \lor 0 \lor -1 \;.
\eea
Therefore we show that the quantization scheme introduced in sec.\ref{sec:foundation} can be reproduced explicitly.

A phenomenological effective action is proposed by Adler\hcite{Adler1} as
\begin{align}
H_{eff} = E_f \sum _i \epsilon ^{\mu\left( i \right)} \gamma _{\mu\left( i \right)} \,, 
\end{align}
where $\epsilon ^\mu $ is the preferred direction and $\left( i \right)$ indicates the site. This action is problematic not only because it has a preferred direction, but also because the propagator of the associated particles would be a tadpole since the action is linear in $X^\mu $. This problem is solved naturally in our theory since we write down the action by directly fermionizing the worldline action with the usual $p^{-2}$ propagator.

\section{Composite Nature and Non-commutativity of Spacetime Measures}
\label{sec:commutator}
Following the same idea presented at the end of sec.III, we can finally define the position operator and the position eigenstate. Since
\bea
\left\langle X^\mu \right\rangle = \sum _i \left\langle e^\mu _I \Delta X^I \right\rangle _{ \left( i \right) } \,,
\eea
one can obtain the form of $\hat{X}^\mu$ and its eigenstate $\left| x \right\rangle$:
\bea
\hat{X}^\mu &=& \lambda \, \sum _i \left( e^\mu _I \gamma ^I _{ \left( i \right) } \underset{j<i}{\otimes} \gamma ^0_{ \left( j \right) } \right) \,,\\
\left| x \right\rangle &=& \underset{i}{\oplus} \left| \Delta x_{ \left( i \right) } \right\rangle \,,
\eea
where $\otimes$ is the direct product, $\underset{i}{\oplus}$ and $\underset{i}{\otimes}$ are the direct sum and the direct product over sites with dummy index $i$. The occurrence of $\gamma ^0$ in the off-site part is purely artificial, and can be removed by choosing other representations, e.g., $\hat{X}^\mu = \sum \lambda \bar{\psi _{ \left( i \right) }} e^\mu _I \gamma ^I \psi _{ \left( i \right) }$. The position operator can be envisioned as the 't HOOFT operator on a 1-$D$ spin chain, and the associated Fourier transformation operator $e^{i k_{\mu}X^{\mu}}$ as the Wilson line operator. This kind of visualization is inherited from the bozonization.

Considering the fact that the commutator of position operators vanishes at large scale, the spacetime interval operators of different sites should commute as long as they are far apart from each other. The simplest assumption would be
\bea
\left[ \gamma ^I_{\left( i \right)} , \gamma ^J_{\left( j \right)} \right] = 0.
\eea
The structure of the spacetime interval operator under this assumption resembles the spin operators in the 4-$d$ Heisenberg model, which is also classical in exterior space but fermionic on site. Notice that the dynamics of the Heisenberg model is completely determined by the Hamiltonian and the commutation relation between nearest neighbors, so does our theory. The assumption we made is therefore equivalent to turning off the non-local dynamics of null geodesic. This makes the structure of the theory greatly simplified, but still provides sufficient insight to the properties of the position operator and the spacetime itself. It is now straightforward to derive the commutation and anti-commutation relations:
\begin{align}
\frac{1}{2 \lambda ^2} \big{[}  \hat{X}^\mu &, \hat{X}^\nu \big{]}  = i \sum _i \left( \sigma ^{\mu\nu}_{\left( i \right)} \underset{j\neq i}{\otimes} I_{\left( j \right)} \right) \,,\\
\frac{1}{2 \lambda ^2} \bigg{\{ } \hat{X}^\mu &, \hat{X}^\nu \bigg{\} } = \sum _i g^{\mu\nu}_{\left( i \right)} \left( \underset{j}{\otimes} I_{\left( j \right)} \right) \nonumber \\
&+ \sum _{i , j\neq i} \left( \gamma ^\mu _{\left( i \right)} \otimes \gamma ^\nu _{\left( j \right)} \underset{i\neq q\neq j}{\otimes} I_{\left( q \right)} \right) \,,\\
\frac{1}{2} \Bigg{\langle} \bigg{\{ } \hat{X}^\mu &, \hat{X}^\nu \bigg{\} } \Bigg{\rangle} = \left\langle \hat{X}^\mu \right\rangle \left\langle \hat{X}^\nu \right\rangle + \sum _i \bigg{(} \lambda ^2 g^{\mu\nu} \nonumber \\
&- \left\langle \Delta X^\mu \right\rangle _{\left( i \right)} \left\langle \Delta X^\nu \right\rangle _{\left( i \right)} \bigg{)} \,,
\end{align}
where $\underset{i\neq q\neq j}{\otimes} I_{\left( q \right)}$ is the direct product over all sites except the $i$-th site and the $j$-th site.
At the large scale limit these relations match the classical results. Non-classical parts of both the anti-commutator and the commutator are of the order of $\lambda X$. The physical quantity corresponding to these corrections are basically the sum of the "surface area" of the discrete spacetime along the curve linking from the origin to the point $X$. It is clear that the spacetime in this work, contrary to those of most other theories, is neither commutative nor anti-commutative.

Another important aspect of non-commutativity of spacetime is the holographic nature of our theory. Since the position eigenstates are composite, there are only $\left(N+1\right)^2$ distinguishable and physically acceptable configurations for states composed by $N$ line elements. A 2-line-element state therefore resembles a composite spin-1 particle, with the proper distance and the area quantum number of the former corresponds to the charge and the spin of the latter, respectively, both with quantum numbers $( 1, 0 ~\rm{or} -1)$. Therefore it is with no doubt that only $\mathcal{O}\left(N^3\right)$ different states are allowed within a 4-sphere, implying that the spacetime information can be written in a 3-dimensional quantum language. With the equation of motion in Eq.(\ref{eom}), the degree of freedom, or entropy, is proportional to surface area of the system.

\section{Generalized Uncertainty Principle and $U(\mathfrak{s}u(2))$ model as semi-classical 3+1 decomposition}
\label{sec:GUP}
In QM, one can define a translation operator $\hat{T}\left(\epsilon\right)$ in terms of the momentum operator:
\bea
\hat{T}_\mu \left( \epsilon \right) &=& e^{-i \epsilon \hat{P}_\mu }\,,
\eea
which satisfies the commutation relations
\bea
\left[ \hat{X}^\mu , \hat{T}_\nu \left( \epsilon \right) \right] &=& \epsilon \, \delta ^\mu _\nu \hat{T}_\mu \left( \epsilon \right) \,,\\
\left[ \hat{T}_\mu , \hat{T}_\nu \right] &=& 0 \,.
\eea
Clearly as long as the inversion of the exponential operator exists, one can reversely construct the momentum operator in terms of the translation operator defined above:
\bea
\hat{P}_\mu &=& i \epsilon \ln \left( \hat{T}_\mu \left( \epsilon \right) \right)  \,, \label{PinT} \\
\left[ \hat{X}^\mu , \hat{P}_\nu \right] &=& \left[ \hat{X}_\mu , i \epsilon \ln \left( \hat{T}_\nu \left( \epsilon \right) \right) \right] \nonumber \\
&=& i \epsilon  \hat{T}_{\mu }(\epsilon ) \frac{\partial \ln \left(\hat{T}_{\nu }(\epsilon )\right)}{\partial \hat{T}_{\mu }(\epsilon )} = i \delta ^\mu _\nu \,,\\
\left[ \hat{P}_\mu , \hat{P}_\nu \right] &=& 0 \,.
\eea
So instead of defining the momentum operator directly, we choose to define the eigenstate of the translation operator, a.k.a. the plane wave solution, for it is less ambiguous, more visualizable, and easier to define.

Since the time direction has nothing to do with our topic of interest (GUP) for the moment, for the sake of simplicity, we will restrict the system to a simpler one with the existence of foliation and use ADM formalism to reduce the dimensions down to three. The system can be understood as a spin chain in the zero temperature limit, with the coordinate measure being the total spin. This system can be reinterpreted as a bosonic system with SU(2) symmetry even at the quantum level, as demonstrated in the non-abelian bosonization. A na\" \i ve guess from QM is that the total spin operator, or the spatial coordinate operator, is now a SU(2) operator of all representations. One can therefore regard this system as exactly the same one in Ref.\cite{SU2F}. As shown in Eqs.(\ref{su2xcommute}) to (\ref{su2dexp}), an immediate conclusion would be the existence of an exterior derivative that can be treated as a momentum operator:
\bea
p_i  &=& dx_i \tilde p_i  \equiv -i dx_i \left( dx_i  \cdot d e^{i k \cdot x} \right) e^{-i k \cdot x} \nonumber \\
     &=& \left( \lambda k \right)^{-1} \sin \left( \lambda k \right) k_i dx_i \,,\\
\left[ x_i , p_j \right] &=& i \delta _{ij} dx_j  + i \lambda \epsilon _{ijk} dx_k \tilde p_j  + \lambda \delta _{ij} d\tau \, \tilde p_j \,,\\
\left[ p_i , p_j \right] &=& i \epsilon _{ijk} dx_k \tilde p_i \tilde p_j \,,\ \ 
\eea
where $p_i$ is the projected momentum that generates translation along $x_i$, and $\tilde p_i$ is its 1-form component.

One can therefore compute the uncertainty relation through
\begin{align}
\Delta x_i \Delta p_j &\geq \frac{1}{2} \Big{[} \left\langle {\left[ {x_i ,p_j } \right]} \right\rangle ^2 \nonumber \\
&+ { \big{(} \left\langle \left\{ x_i , p_j \right\} \right\rangle - 2 \left\langle x_i \right\rangle \left\langle p_j \right\rangle \big{)}^2 \Big{]} }^{1/2} \,.
\end{align}

In the low momentum limit ($\left\langle k \right\rangle$ small and $\left\langle x \right\rangle \to 0$), we obtain
\bea
\Delta x_i \Delta p_j  \geqslant \frac{1}{2} \left( \delta _{ij} + 2 \lambda ^2 \delta _{ij} \left\langle k_i \right\rangle \left\langle k_j \right\rangle \right) \,.
\eea
The GUP of our theory is of the usual form. (e.g., Snyder's GUP in Eq.(\ref{snyderGUP}) is identical to ours with relation $\lambda = \sqrt{2} a$)

%\section{Emergence of Lorentzian Manifold and Holographic Principle at Large Scale Limit}
%\label{sec:holographic}
%Assuming a random distribution, one can derive a "Fermi-Dirac probability distribution" for quantum spacetime measure
%\bea
%P_k &=& \frac{1}{1+e^{\beta\left(A_k-A_0\right)}} \,,\\
%n &=& \int \epsilon _k P_k dk \,,\\
%A &=& \int A_k \epsilon _k P_k dk \,,\\
%R &=& \frac{A}{n} \,,
%\eea
%where $A_k$ is the area element of the state $k$ ($A_k \approx k$ in our case) ,$A_0$ is a constant area that resembles Fermi level, $\beta$ is a function of the constraint $A$ (resembles $\beta$ and total energy in thermodynamics), and $\epsilon _k$ is the density of state ($\epsilon _k =k$ in our case). Since distance measure is exactly the same as area element measure, we have average distance from the surface $R=A/n$.

%Assuming classical limit
%\bea
%P_k &\approx & e^{-\beta A_k} / Z \,,\\
%Z &=& \int \epsilon _k e^{-\beta A_k} dk \,,
%\eea
%where Z is the normalizing factor, one can derive Bekenstein's law by assuming $A=4\pi R^2$
%\bea
%\beta &=&\sqrt{\frac{4\pi}{A}} \,,\\
%S &=& \frac{ \left( \pi ^2+6 \zeta \left( 3\right)\right)}{24\pi} A \approx \frac{0.906}{4}A \,.
%\eea

\section{Astrophysical Test}
\label{sec:astrophystest}
It is straightforward to apply the quantized geodesic obtained in Sec.\ref{sec:commutator} to trajectories of cosmic rays, which may serve as an astrophysical test for this theory through the smearing of particle trajectory. Consider, for example, the null geodesic of a 10 MeV photons emitted from a gamma ray burst afterglow at high redshift ($z \approx 10$). Because the Lorentz invariance is manifest in our theory, the smearing effect occurs only along the perpendicular direction. Assuming Gaussian distribution of photons at source, we find, from Eq.(\ref{var_dX}) and Eq.(\ref{var_ds}), that the smearing of the photon is of the order of $\sqrt{ E \lambda ^2 \; d_S \left( z \approx 10 \right) } \approx {10}^{-14} \lambda$ meter, where $d_S = \int _0^z \frac{(1+z')dz'}{H(z')}$ is the rescaled distance taking the rescaling of the variance into account, $E$ is the GRB characteristic energy, and $\lambda$ is the scale of our theory in the unit of the Planck length. Clearly it's impossible to identity such tiny effect in the near future.

Let us consider an EeV neutrino at high redshift ($z \approx 10$) as another example. The smearing for massive particles along the perpendicular direction is of the order of $\sqrt{ \frac{E}{m_\nu} \lambda \; d_S \left( z \approx 10 \right) } \approx 100 \sqrt{\lambda}$ meter. For time-like geodesic, in addition to perpendicular direction, there are also variance in proper distance, which allow us for more precise measurement. The smearing is at most $\sqrt{3 \lambda t_P \; T \left( z \approx 10 \right) } \approx \sqrt{\lambda}$ picosecond, where $t_P$ is the Planck time and $T$ is the time taken from $z=10$ to $z=0$. One possible way to probe the smearing of the arrival time is by observing both neutrinos and gravitational waves from a single merger event of the black hole-neutron star binary. Since typically the neutrino flux bumps up sharply 0.5 ms after the merger event\hcite{neutronmerger}, a time-resolution of 0.5 ms can be achieved by detecting neutrinos right after the detection of the gravitational merger event. The resolution of $\lambda$ is therefore of the order $10^{-17}$ meter. An improvement on the simulation could possibly enhance the time-resolution in this merger scenario and allow a tighter constraint on the value of $\lambda$.

\section{Conclusion and Future Work}
\label{sec:conclusion}

We show that a non-commutative spacetime theory can be constructed on top of Adler's ``linear line element", and derive the associated action from the fermionization of the K-K ground state of the Polyakov action on a cylinder. The theory is shown to be holographic, and passes the astrophysical tests on the smearing effect and Lorentz invariance violation. The theory can also be regarded as the $4$-D extension to the $U(\mathfrak{s}u(2))$ model, and from there the generalized uncertainty principle is derived.

The theory is quite potent, and many aspects of the theory still remain unsolved. E.g., one may try to fix the Bekenstein-Hawking coefficient k inside Bekenstein's formula $S = k A / 4$. We have also introduced kinematics of the theory only. The dynamics of the theory can be introduced through adding loop quantum gravity to the theory, or by adding back the K-K tower of the worldsheet excitation.
Interestingly in loop quantum gravity\hcite{LQGlength2}, it has been argued that the length operator should be defined through Dirac operators. It remains curious whether our work can be served as a representation that can describe length, area and volume easily without introducing the spectral decomposition\hcite{LQGlength0,LQGlength1}.

\section{Acknowledgement}
We appreciate useful discussions with Ron Adler and Abhay Ashtekar. This work is supported by National Center for Theoretical Sciences (NCTS) of Taiwan, Ministry of Science and Technology (MOST) of Taiwan, and the Leung Center for Cosmology and Particle Astrophysics (LeCosPA) of National Taiwan University.

%\begin{appendices}
%\section{Structure of Local Coordinate Difference Operator}
%\label{sec:Append-operator}

%\section{Discretization of Distance, Area and Volume}
%\label{sec:Append-Eigenvalues}

%\end{appendices}


\begin{thebibliography}{99}
  
\bibitem{LorentzInvExp}
  David~Mattingly,
   ``Modern Tests of Lorentz Invariance'',
   {\it Living Rev. Relativity\,} {\bf 8}, 5 (2005),
   \href{http://arxiv.org/abs/gr-qc/0502097}{arxiv:gr-qc/0502097}.

\bibitem{Snyder}
  Hartland~S.~Snyder,
   ``Quantized Space-Time'',
  \href{http://journals.aps.org/pr/abstract/10.1103/PhysRev.71.38}{{\PR} {\bf 71}, 38 (1947)}.

\bibitem{Bicrossproduct}
  Shahn~Majid and Henri~Ruegg,
   ``Bicrossproduct structure of $\kappa$-Poincar\' e group and non-commutative geometry'',
   {\PL} {\bf B334}, 348 (1994),
   \href{http://arxiv.org/abs/hep-th/9405107}{arxiv:hep-th/9405107}.

\bibitem{DSRBasis}
  Jerzy~Kowalski-Glikman and Sebastian~Nowak,
   ``Non-commutative space-time of Doubly Special Relativity theories'',
   {\IJMP} {\bf D12}, 299 (2003),
   \href{http://arxiv.org/abs/hep-th/0204245}{arxiv:hep-th/0204245}.

\bibitem{DSRIntro}
  Jerzy~Kowalski-Glikman,
   ``Introduction to Doubly Special Relativity'',
   {\it Planck Scale Effects in Astrophysics and Cosmology} pp 131, Springer Berlin Heidelberg (2005),
   \href{http://arxiv.org/abs/hep-th/0405273}{arxiv:hep-th/0405273}.

%\bibitem{DSRDispersion}
%  Giovanni~Amelino-Camelia and Shahn~Majid,
%   ``Waves on Noncommutative Spacetime and Gamma-Ray Bursts'',
%   {\IJMP} {\bf A15}, 4301 (2000),
%   \href{http://arxiv.org/abs/hep-th/9907110}{arxiv:hep-th/9907110}.

\bibitem{Connes1}
  Alain~Connes and John~Lott,
   ``Particle models and noncommutative geometry'',
   {\NP} {\bf B18}, 29 (1991),
   \href{http://deepblue.lib.umich.edu/bitstream/handle/2027.42/29524/0000611.pdf}{library of UMichigan}.

\bibitem{Connes2}
  Alain~Connes,
   ``Gravity coupled with matter and the foundation of non commutative geometry'',
   {\CMP} {\bf 182}, 155 (1996),
   \href{http://arxiv.org/abs/hep-th/9603053}{arxiv:hep-th/9603053}.

\bibitem{SU2F}
  Eliezer~Batista and Shahn~Majid,
   ``Noncommutative Geometry Of Angular Momentum Space $U(\mathfrak{s}u(2))$'',
   {\JMP} {\bf 44}, 107 (2003),
   \href{http://arxiv.org/abs/hep-th/0205128}{arxiv:hep-th/0205128}.

\bibitem{SL2CF}
  Laurent~Freidel and Shahn~Majid,
   ``Noncommutative harmonic analysis, sampling theory and the Duflo map in 2+1 quantum gravity'',
   {\CQG} {\bf 25}, 045006 (2008),
   \href{http://arxiv.org/abs/hep-th/0601004}{arxiv:hep-th/0601004}.

\bibitem{3DQG}
  Eliezer~Batista and Shahn~Majid,
   ``Three dimensional quantum geometry and deformed symmetry'',
   {\JMP} {\bf 50}, 052503 (2009),
   \href{http://arxiv.org/abs/0806.4121}{arxiv:0806.4121}.

\bibitem{ConstNoncomm}
  Sergio~Doplicher, Klaus~Fredenhagen, and John~E.~Roberts,
   ``The quantum structure of spacetime at the Planck scale and quantum fields'',
   {\CMP} {\bf 172}, 187 (1995),
   \href{http://arxiv.org/abs/hep-th/0303037}{arxiv:hep-th/0303037}.

\bibitem{Adler1}
  Ronald~J.~Adler,
   ``A quantum theory of distance along a curve'',
   Unpublished (2014),
   \href{http://arxiv.org/abs/1402.5921}{arxiv:1402.5921}.

%\bibitem{Spin4}
%  Jungjai~Lee, John~J.~Oh, and Hyun~Seok~Yang,
%   ``An Efficient Representation of Euclidean Gravity I'',
%   {\JHEP} {\bf 12}, 025 (2011),
%   \href{http://arxiv.org/abs/1109.6644}{arxiv:1109.6644}.

\bibitem{qdeform}
  U.~Carow-Watamura, M.~Schlieker, M.~Scholl and S.~Watamura,
   ``Tensor Representation of the Quantum Group $SL_q (2, \mathbb{C})$ and Quantum Minkowski Space'',
   {\ZP} {\bf C 48}, 159 (1990),
   \href{http://link.springer.com/article/10.1007/BF01565619}{Springer}.

%\bibitem{NoncommString}
%  Nathan~Seiberg and Edward~Witten,
%   ``String Theory and Noncommutative Geometry'',
%   {\JHEP} {\bf 09}, 032 (1999),
%   \href{http://arxiv.org/abs/hep-th/9908142}{arxiv:hep-th/9908142}.

%\bibitem{NonCommQFT}
%  Michael~R.~Douglas and Nikita~A.~Nekrasov,
%   ``Noncommutative field theory'',
%   {\RMP} {\bf 73}, 977 (2001),
%   \href{http://arxiv.org/abs/hep-th/0106048}{arxiv:hep-th/0106048}.

%\bibitem{QFTNonCommSpace}
%  Richard~J.~Szabo,
%   ``Quantum field theory on noncommutative spaces'',
%   {\PRep} {\bf 378}, 207 (2003),
%   \href{http://arxiv.org/abs/hep-th/0109162}{arxiv:hep-th/0109162}.

%\bibitem{GravNonComm}
%  Richard~J~Szabo,
%   ``Symmetry, gravity and noncommutativity'',
%   {\CQG} {\bf 23}, R199 (2006),
%   \href{http://arxiv.org/abs/hep-th/0606233}{arxiv:hep-th/0606233}.

%\bibitem{LQGBH}
%  Krzysztof~A.~Meissner,
%   ``Black hole entropy in Loop Quantum Gravity'',
%   {\CQG} {\bf 21}, 5245 (2004),
 %  \href{http://arxiv.org/abs/gr-qc/0407052}{arxiv:gr-qc/0407052}.

\bibitem{LQGlength0}
  T.~Thiemann,
   ``A length operator for canonical quantum gravity'',
   {\JMP} {\bf 39}, 3372-3392 (1998),
   \href{http://arxiv.org/abs/0806.4710}{arxiv:gr-qc/9606092}.

\bibitem{Peskin}
  M. Peskin and D. Schroeder, {\it Introduction to Quantum Field Theory}

\bibitem{bosonization}
  Sidney~Coleman,
   ``Quantum sine-Gordon equation as the massive Thirring model'',
   {\PR} {\bf D11}, 2088 (1975),
   \href{http://users.physik.fu-berlin.de/~kamecke/ps/coleman.pdf}{Freie Universit\"at Berlin}.

\bibitem{nonabelianbosonization}
  Edward~Witten,
   ``Non-Abelian Bosonization in Two Dimensions'',
   {\CMP} {\bf 92}, 455-472 (1984),
   \href{https://projecteuclid.org/euclid.cmp/1103940923}{Project Euclid}.

\bibitem{Steinhardt}
  Paul~J.~Steinhardt,
   ``Baryons and baryonium in QCD${}_2$'',
   \href{http://www.sciencedirect.com/science/article/pii/0550321380900656}{{\NP} {\bf B176}, 100-112 (1980)}.

%\bibitem{neutrinoLIV}
%  Enrico~Borriello, Sovan~Chakraborty, Alessandro~Mirizzi, and Pasquale~Dario~Serpico,
%   ``Stringent constraint on neutrino Lorentz invariance violation from the two IceCube PeV neutrinos'',
%   {\PR} {\bf D87}, 116009 (2013),
%   \href{http://arxiv.org/abs/1303.5843}{arxiv:1303.5843}.

\bibitem{neutronmerger}
  Francois~Foucart, Evan~O'Connor, Luke~Roberts, Matthew~D.~Duez, Roland~Haas, Lawrence~E.~Kidder, Christian~D.~Ott, Harald~P.~Pfeiffer, Mark~A.~Scheel, and Bela~Szilagyi
   ``Post-merger evolution of a neutron star-black hole binary with neutrino transport'',
   {\PR} {\bf D91}, 124021 (2015),
   \href{http://arxiv.org/abs/1502.04146}{arxiv:1502.04146}.

\bibitem{LQGlength2}
  Carlo~Rovelli,
   ``Lorentzian Connes Distance, Spectral Graph Distance and Loop Gravity'',
   Unpublished (2014),
   \href{http://arxiv.org/abs/1408.3260}{arxiv:1408.3260}.

\bibitem{LQGlength1}
  Eugenio~Bianchi,
   ``The length operator in Loop Quantum Gravity'',
   {\NP} {\bf B807}, 591 (2009),
   \href{http://arxiv.org/abs/0806.4710}{arxiv:0806.4710}.

\end{thebibliography}
\end{document}